# Three Photon Absorption in Optical Parametric Oscillators Based on OP-GaAs

OLIVER H. HECKL,[1*] BRYCE J. BJORK,[1] GEORG WINKLER,[2] P. BRYAN CHANGALA,[1] BEN SPAUN,[1] GIL PORAT,[1] THINH Q. BUI,[1] KEVIN F. LEE,[3] JIE JIANG,[3] MARTIN FERMANN,[3] PETER G. SCHUNEMANN[4] AND JUN YE[1*]

[1]JILA, National Institute of Standards and Technology and University of Colorado, Department of Physics, University of Colorado, Boulder, CO 80309, USA
[2]Institute of Atomic and Subatomic Physics, Vienna University of Technology, Stadionalle 2, 1020 Vienna, Austria
[3]IMRA America, Inc., 1044 Woodridge Ave. Ann Arbor Michigan 48105, USA
[4]BAE Systems, PO Box 868 Nashua New Hampshire 03063, USA

*Corresponding author: oliver.heckl@jila.colorado.edu

**We report on the first singly resonant (SR), synchronously pumped optical parametric oscillator (OPO) based on orientation-patterned gallium arsenide (OP-GaAs). Together with a doubly resonant (DR) degenerate OPO based on the same OP-GaAs material, the output spectra cover 3 to 6 μm within ~3 dB of relative power. The DR-OPO has the highest output power reported to date from a femtosecond, synchronously pumped OPO based on OP-GaAs. We discovered strong three-photon absorption with a coefficient of 0.35 ± 0.06 cm$^3$/GW$^2$ for our OP-GaAs sample, which limits the output power of these OPOs as mid-IR light sources. We present a detailed study of the three-photon loss on the performance of both the SR and DR-OPOs, and compare them to those without this loss mechanism.**

OCIS codes: (190.4360) Nonlinear optics, devices; (190.7110) Ultrafast nonlinear optics; (190.4400) Nonlinear optics, materials; (190.4970) Parametric oscillators and amplifiers.

http://dx.doi.org/10.1364/OL.99.099999

Mid-infrared frequency combs have emerged as a new generation of spectroscopic tools for molecular science [1–3] due to their ability to quickly acquire a broadband molecular spectrum with high spectral resolution. High average power frequency combs allow for increased sensitivity and shorter acquisition times. To date, the highest power frequency comb source in the mid-IR beyond 3 microns is a singly resonant optical parametric oscillator (SR-OPO) based on a lithium niobate crystal (transparent out to 5 μm), with >1 W average output power [1]. However, frequency comb techniques are currently limited in the fingerprint region beyond 5 microns where no watt-level high-power sources exist [4–6]. Gallium arsenide (GaAs) is considered a promising material for the generation of mid-IR in the fingerprint region due to its large transparency range from 0.9 to 17 μm (bandgap of 1.42 eV) [7,8], its large nonlinear coefficient of $d_{14}$ = 94 pm/V (at 4 μm) [9], and the availability of quasi-phase matching (QPM).

In this Letter we demonstrate the first SR-OPO based on OP-GaAs. We were able to achieve a maximum output power of 11 mW around 5.5 μm at a threshold pump power of 940 mW. The SR-OPO is built around a 470 μm thick OP-GaAs crystal with a QPM-period of 58 μm. The crystal was mounted at Brewster angle and rotated by 35° to the [011] crystallographic direction. This assures the highest nonlinear coefficient along [111] as well as linearly p-polarized light in the signal and idler [9].

A detailed schematic of the SR-OPO is shown in Fig. 1. The resonator is set up as a standing wave cavity which drastically simplifies alignment, at the expense of increased intracavity loss. One of the flat mirrors is used as an input coupler for the pump, and the other flat mirror serves as an output coupler for the generated idler. Both flat mirrors are highly transparent for the pump laser operating at 1.95 μm. The pump laser is a Tm-fiber laser that is amplified to 2 W average output power. The laser has a repetition rate of 110 MHz and a pulse duration of 157 fs.

With the pump power threshold, $P_{th}$, of 940 mW and the available pump power of 1.8 W inside the resonator, our expectation was to generate at least several tens of mW in the mid-IR, as shown by the red dashed line in Fig. 2a. However, we found a strong saturation of the output power, as evidenced by the measurement (blue circles) in Fig. 2a.

We identified three-photon absorption (3PA) as a limiting factor for the performance of optical parametric oscillators based on GaAs. Previous measurements of the 3PA coefficient γ yielded a value of

0.35 ± 0.5 cm³/GW² at a wavelength of 2.3 μm and a value of 0.25 cm³/GW² at 2 μm [10].

In order to characterize the exact contribution of 3PA in our system, we measured the 3PA coefficient γ of the OP-GaAs crystal in our setup. The z-scan measurements were carried out according to [11]. The crystal was placed close to perpendicular to the incident beam and the polarization had an angle of 35° to the [011] crystallographic direction. The pulse duration of the amplified mode-locked Tm-fiber laser was measured with the frequency-resolved optical gating (FROG) technique. The wavelength of the laser is centered around 1.95 μm, which corresponds to 44% of the bandgap of GaAs. The spot size in the focus was 15.6 ± 0.4 μm and was determined from the fit parameters of the z-scan. This leads to a peak on-axis intensity of 10 ± 1 GW/cm². Since the OP-GaAs crystal was not anti-reflection coated we corrected the incident power for the Fresnel reflection on the surface. A z-scan measurement in the patterned region of the GaAs crystal is shown in Fig. 3. We analyzed the z-scan data according to [10,11] and determined that 3PA is indeed the dominant absorption mechanism at our pump wavelength of 1.95 μm.

We measure the three photon coefficient in our orientation-patterned GaAs crystal (58 μm pattern) to be 0.30 ± 0.06 cm³/GW² in the substrate, and 0.35 ± 0.06 cm³/GW² in the patterned region of the crystal. The increased values for the three photon absorption coefficient in the patterned region might be due to the existence of mid-gap states in GaAs [12] or an increased free carrier lifetime in hydride vapor phase epitaxy grown samples due to the absence of defect states which decrease carrier lifetimes [13].

The three photon absorption coefficient of GaAs is high compared to, for example, the three photon absorption coefficient of ZnO or Si, reported with values of 0.003 ± 0.0015 cm³/GW² [11] and 0.022 ± 0.01 cm³/GW² [14], respectively. Both values were measured with photons carrying an energy of 44% of their respective bandgap (3.3 eV for ZnO and 1.1 eV for Si).

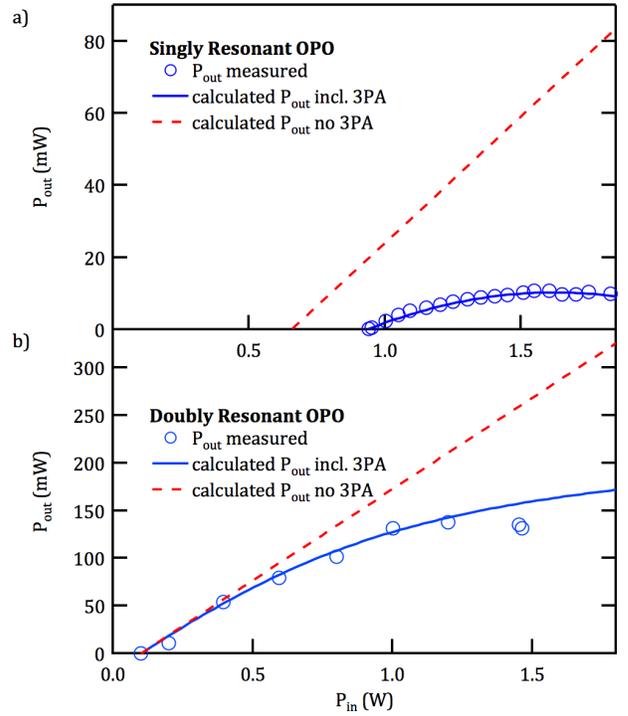

Fig. 2: Output power versus pump power for the SR-OPO (a) and the DR-OPO (b). The red dashed lines indicate the output power of the OPOs without 3PA. The blue circles are the measured output power and the blue solid line is the calculated output power based on our measurements for the 3PA coefficient and the threshold of the OPOs

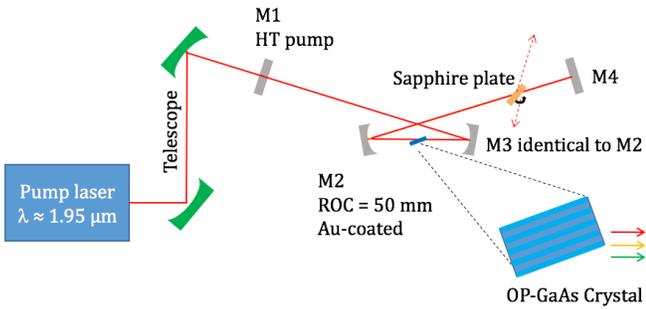

Fig. 1: Setup of the SR- and DR-OPO. A telescope matches a 1.95 μm, 110 MHz modelocked laser to a standing wave cavity formed by M1-4. An orientation-patterned gallium arsenide (OP-GaAs) crystal is placed at the cavity focus between M2 and M3, with a waist size of 21 μm. **SR-OPO:** M1 and M4 are HT for the pump and HR for the signal wavelength (2.5 – 3.1 μm). M4 output couples the idler wavelength (T~>85%). **DR-OPO:** M1 is HR at 3.0 – 6.6 μm with 80% pump transmission and M4 is a broadband gold mirror. Light is coupled out of the OPO at the sapphire Brewster plate.

Losses from 3PA have a quadratic dependence on input power. These additional losses will increase the threshold of the OPO. Their influence on the threshold of the OPO can be calculated with the threshold criteria for SR-OPOs under the assumption of perfect phase matching [15,16],

$$g^2 L_{eff}^2 = 2l = 2(l_0 + l_{3PA}). \quad (1)$$

$L_{eff}$ stands for the effective crystal length, $l_0$ denotes the losses for the signal in the OPO without 3PA and $l_{3PA}$ is the induced loss due to 3PA. The parametric gain coefficient, $g$, links the losses to the average pump power, $P_{av}$, and is given by the expression [15]

$$g^2 = \frac{8\pi^2 d_{eff}^2 P_{av}}{n_p n_s n_i \lambda_s \lambda_i c \epsilon_0 f_{rep} \tau_p w_0^2 \pi} \frac{\tau_p^2}{\tau_p^2 + \tau_S^2}. \quad (2)$$

Here we assume perfect spatial coupling. $n_p, n_s, n_i$ refer to the refractive indices of the pump, signal and idler. $\lambda_s$ and $\lambda_i$ refer to their respective wavelengths. $\tau_p$ and $\tau_s$ correspond to the pulse duration of the pump laser and the signal, respectively.

The output power of a SR-OPO can then be expressed as

$$P_{out,3PA} = D_{SR,i}(P_{in} - P_{th,3PA})$$
$$= D_{SR,i}(P_{in} - P_{th,no\ 3PA} - bP_{in}^2) \quad (3)$$

where $D_{SR,i}$ is the photon conversion efficiency into the idler, $b$ is calculated via equation (1) and (2) including the measured 3PA loss coefficient. $b$ has a value of 0.31 W⁻¹ in our system. Clearly, $P_{th,3PA}$ is input power dependent.

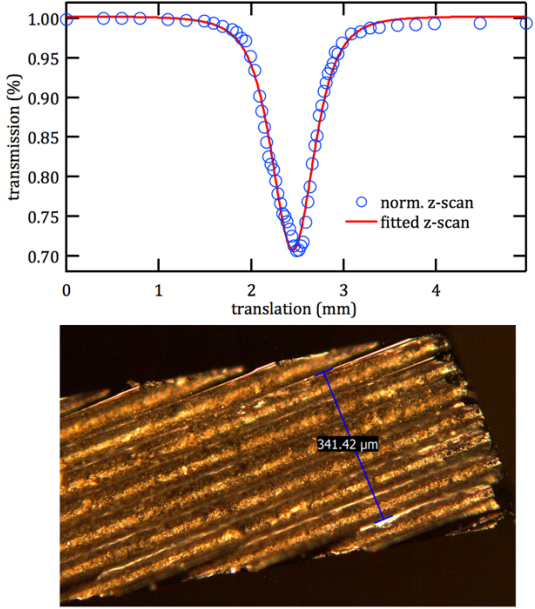

Fig. 3: Top: Z-scan measured in the patterned region of a 58 μm QPM-period OP-GaAs crystal. This measurement reveals a three photon absorption coefficient of 0.35 ± 0.06 cm3/GW2. Similar measurements in the substrate (non-patterned region) showed a low

For the SR-OPO, we estimate the losses $l_0$ to be 10.2%, limited mainly by the mirrors used for the OPO cavity. At the threshold of the SR-OPO we calculate a value of 4.2% for $l_{3PA}$. Here we assume that the 3PA loss (caused by the pump light) carries over to the signal. We used $\tau_s$ as a fitting parameter to self-consistently describe the losses at threshold and the threshold power, $P_{th,3PA}$, of 940 mW. The signal pulse duration is then kept constant. In a second step, we calculate the threshold without 3PA, $P_{th, no\ 3PA}$ to be 660 mW.

The conversion efficiency $D_{SR,i}$ was adjusted in order to achieve the best fit of the theoretical model to our measured output power. In this way we find a total photon conversion efficiency of $D_{SR}$ of 20%, This corresponds to a conversion efficiency into the idler $D_{SR,i}$ of 7%.

The measured output power versus pump power can be seen in Fig. 2a. The output power has a maximum of 11 mW and follows a quadratic behavior that is well described by Eq. (3). An idealized SR-OPO with the slope efficiency $D_{SR}$ and without 3PA losses is displayed to highlight the detrimental effect of 3PA in Fig. 2a. An output power of about 90 mW in the idler is calculated in the absence of 3PA, even at our relatively low photon conversion efficiency of 20%. The non-degenerate spectra of signal and idler of the SR-OPO are displayed in Fig. 4. No effort was undertaken to optimize the intracavity dispersion and hence the spectral coverage of the singly resonant OPO as has been demonstrated for the case of a DR-OPO [17,18].

After we had understood the limiting factor of 3PA for the SR-OPO, we exchanged the in-coupling and out-coupling mirrors of the OPO to enable doubly resonant degenerate operation. To enable direct comparison of the two OPOs we kept the mode matching, crystal position and cavity type identical to the SR case. We exchanged the OP-GaAs crystal for a 510 μm thick, 53 μm QPM period, in order to allow for degenerate operation of the DR-OPO.

As in the SR case the crystal was mounted at Brewster angle and rotated by 35° with respect to the [011] crystallographic direction.

The pump is coupled into the cavity via a flat input coupling mirror, M1 in Fig. 1, with about 80% pump transmission and high reflectivity from 3.0 – 6.6 μm, supporting the operation of a degenerate DR-OPO. The remaining mirror, M4, is chosen to be another broadband gold coated mirror. We used a sapphire plate on a rotation stage as an output coupler of variable reflectivity.

The output power, $P_{out}$, as a function of pump power, $P_{in}$, of the DR-OPO is displayed in Fig. 2b. The DR-OPO was optimized for output power while maintaining the mode matching and spot size of the SR-OPO by gradually increasing the output coupling losses $\eta_{OC}$ to 16.7%. At this relatively high outcoupling-rate we estimate a total loss of about $l_{tot,th}$ = 28% for the signal / idler of the OPO at the threshold of operation. We achieved a maximum output power of 137 mW (summed over all four reflections from the sapphire plate) at a pump power of 1.2 W measured inside the cavity after the input coupling mirror. The output power of the DR-OPO clearly deviates from the expected linear slope [16], see Fig. 2b.

The linear slope (red dashed line) is the expected output power $P_{out, no\ 3PA}$ without the inclusion of any power-dependent losses. Its zero-crossing is given by the measured threshold, $P_{th}$, of the DR-OPO of 100 mW. The slope of this idealized output power is given by photon conversion efficiency, $D_{DR}$, of 30% (deduced from the measured pump depletion) and the outcoupling rate, $\eta_{OC}$, in relation to the total losses, $l_{tot,th}$, and has the form

$$P_{out,no\ 3PA} = \frac{D\eta_{OC}}{l_{tot,th}}(P_{in} - P_{th}). \quad (4)$$

Analogously to the SR-OPO, we include the effect of 3PA on the performance of the DR-OPO by first calculating the pump power dependent losses that are introduced into the system. Here we use our measured value for 3PA in OP-GaAs.

The influence of 3PA on the threshold of the DR-OPO can be calculated with the threshold criteria for DR-OPOs under the assumption of perfect phase matching [15,16],

$$g^2 L_{eff}^2 = l^2. \quad (5)$$

In the case of the DR-OPO, the additional losses due to 3PA will also decrease the outcoupling rate to $\eta_{OC}/(l_{tot} + l_{3PA})$, see Eq. (4).

For the DR-OPO the mirror M1 (see Fig. 1) had a pump reflectivity of 20%. The remaining mirrors M2-M4 were all gold coated. This led to a slight pump enhancement within the cavity as well as a second pass of the high average power pump light through the GaAs crystal. As in the SR-OPO, the additional pump power dependent losses, $l_{3PA}$, led to an increased threshold of the DR-OPO, $P_{th,3PA}$. Considering the above effects, the output power $P_{out,3PA}$ of the DR-OPO can be written as

$$P_{out,3PA} = \frac{D_{DR}\eta_{OC}}{l_{tot} + l_{3PA}}(P_{in} - P_{th,3PA}). \quad (6)$$

The measured output power versus pump power is displayed in Fig. 2b, blue open circles. The prediction of Eq. (6) is depicted with a solid blue line. Inclusion of 3PA effects provides good agreement with the measured performance of the DR-OPO.

Instead of a linear increase in output power, we observed saturation behavior due to the increase in losses and the decrease

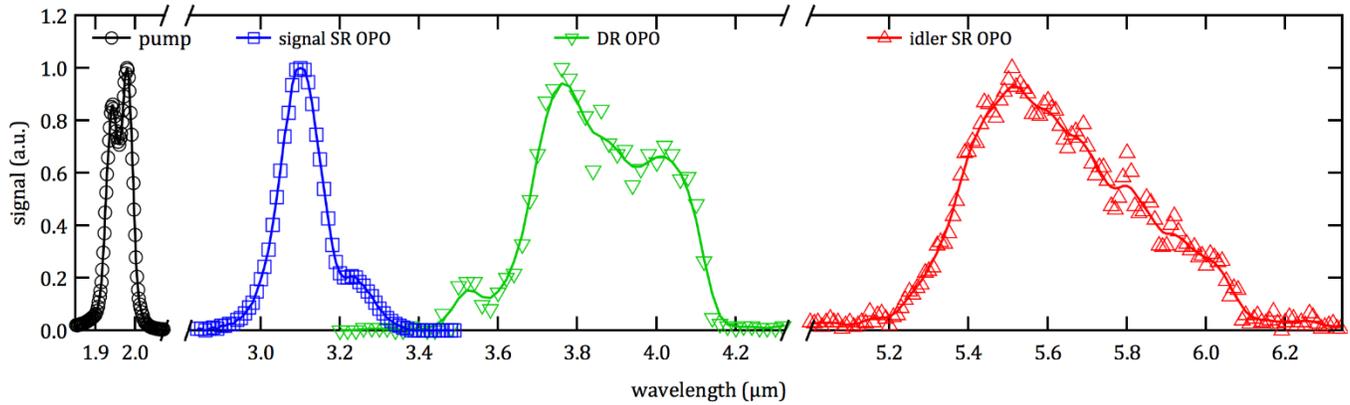

Fig. 4: Spectra of the pump laser (black circles), the output of doubly resonant OPO (green, inverted pyramids), signal of singly resonant OPO (blue squares) and the idler of the singly resonant OPO (red pyramids). The spectrum of the pump laser was measured with a commercial mid-IR spectrometer. The remaining spectra were measured with a monochromator and a mid-IR photodiode

in the relative outcoupling rate. Nevertheless, to the best of our knowledge, the reported output power here is the highest output power ever reported from a femtosecond, synchronously-pumped DR-OPO based on OP-GaAs [4,5,18]. In the work of V. Smolski et al. [4,18], 3PA is not apparent due to the lower available pump power. In comparison, the output power of our DR-OPO could be further increased by using a ring cavity configuration and hence reducing the cavity roundtrip losses and threshold. At our highest available pump power (measured within the OPO-cavity) of 1.46 W the losses for the signal / idler wavelength increase from 28% to 41.5%, which reduces the outcoupling rate from 63% down to 40%.

The spectrum of the DR-OPO is displayed in Fig. 4. The degenerate DR-OPO produces a spectrum that spans 600 nm around a central wavelength of 3.9 µm. No effort was undertaken to optimize the intracavity dispersion, but this would increase spectral coverage [17,18].

In this Letter we have presented what is to our knowledge the first SR-OPO based on OP-GaAs. In addition, we have presented the highest output power of a DR-OPO based on the same nonlinear material. The output power of the DR-OPO could be further increased by working with a larger mode size and a ring cavity.

The performance of both the SR- and DR-OPO in high peak power regimes are greatly hindered by the presence of 3PA in GaAs if pumped with wavelengths with photon energies corresponding to a third of its bandgap or more. In particular, SR-OPOs are very susceptible to intracavity losses. In our case, this meant that the SR-OPO based on OP-GaAs barely met the threshold condition. It did, however, allow us to clearly model and understand its limitations in light of 3PA.

Our results show that for future work on high average power mid-IR light sources in the fingerprint region beyond 5 µm, significant OP-GaAs process improvements will be needed to minimize extrinsic defect contributions to 3PA when pumped with standard wavelengths (Yb- or Tm based laser systems). Our results also highlight the benefits of gallium phosphide (GaP), a promising new alternative with a higher 2.26 eV bandgap that also allows for QPM. Recently, a study of SR-OPOs based on GaP has been demonstrated by Maidment et al. [19] where the authors extended wavelength coverage to above 11 µm at several mW of output power. The system was pumped with a 1 µm source of 2 W average power.

**Acknowledgements:** We would like to thank V. O. Smolski and K. L. Vodopyanov for helpful discussions. We acknowledge funding from DARPA SCOUT, AFOSR, NIST, and NSF-JILA PFC for this research. OHH is partially supported through an Alexander-von-Humboldt Fellowship. PBC is supported by the NSF GRFP (Award No. DGE1144083). BS and TQB are supported through the NRC Postdoctoral Fellowship program.